# Quantum Thermodynamics and Quantum Coherence Engines

## Aslı TUNCER[1]*, Özgür E. MÜSTECAPLIOĞLU[1]
[1]Department of Physics, Science Faculty, Koc University, Istanbul, Turkey

**Abstract:** Advantages of quantum effects in several technologies, such as computation and communication, have already been well appreciated, and some devices, such as quantum computers and communication links, exhibiting superiority to their classical counterparts have been demonstrated. The close relationship between information and energy motivates us to explore if similar quantum benefits can be found in energy technologies. Investigation of performance limits for a broader class of information-energy machines is the subject of the rapidly emerging field of quantum thermodynamics. Extension of classical thermodynamical laws to the quantum realm is far from trivial. This short review presents some of the recent efforts in this fundamental direction and focuses on quantum heat engines and their efficiency bounds when harnessing energy from non-thermal resources, specifically those containing quantum coherence and correlations.

## 1. Introduction

Nowadays, we enjoy an unprecedented degree of control on physical systems at the quantum level, which promises exciting developments of machines, such as quantum computers, which can be supreme to their classical counterparts [1]. Learning from the history of industrial revolutions (IRs) associated with game-changing inventions such as steam engines, lasers, computers, the natural question to ask what are the fundamental limits of those promised machines of quantum future.

Intriguingly, every IR can be associated with an invention of a "heat engine". The first IR involves harnessing mechanical energy with signature developments in textile production using the steam engine. Electrical energy fueled the second IR, yielded progress in mass production with electrical motors. Signature machines of that era were based upon Otto and diesel heat engines, converting heat into mechanical work. In terms of computational and laser-based devices, the power of information and radiation is central in the third IR. Preceding invention of laser, known as maser, operates in principle as an Otto engine, too [2]. Surprisingly, despite their dramatic progress, neither of such devices, ranging from ancient machines, such as Hero's aeolipile, to the modern ones, such as masers [3, 4], can beat the efficiency bound established in 1824 by Nicolas- Léonard-Sadi Carnot [5].

His father named sadi Carnot after the pen name of the celebrated Persian poet and philosopher, Saadi of Shiraz (Muslih al-Din). The second law of thermodynamics has a poetic characteristic as well; different people may interpret it differently. Carnot has discovered a practical version of it, setting a universal bound on the operation of engines based upon the motive power of heat. He stated that no heat engine could be more efficient than a specific limit. It is determined only by the temperatures of the hot and cold heat baths and independent of any unique mechanism or any particular agent. His efforts transformed the art of heat engine design to the science of thermal machines, a foundation pillar of thermodynamics. Due to his premature death from the cholera epidemic in 1832, and burial of his unpublished works with him, we do not know if Sadi Carnot could have more results on the laws of thermodynamics. Still, his Carnot bound contribution is more than enough to bring his name among the founding fathers of the thermodynamics.

---

*Correspondence: asozdemir@ku.edu.tr

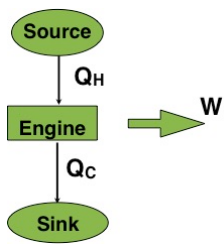

Figure 1: Simple model for a heat engine. $Q_H$ is input heat from a hot source to the engine, $Q_C$ is the heat rejected to the colder environment (entropy sink) and $W$ is the work output.

Carnot engine is an exceptional yet too ideal engine. It operates over a completely reversible, frictionless cycle of isothermal heating-cooling and adiabatic expansion-compression steps. While such a cycle becomes independent of implementations and gives maximum efficiency (ratio of the net output work to the input heat) of

$$\eta_C = 1 - \frac{T_C}{T_H} \tag{1.1}$$

that one can achieve between a hot bath and a cold bath at temperatures $T_C$ and $T_H$, respectively, it produces no power due to its infinitely slow operation. Any engine operating between two heat reservoirs cannot be more efficient than a Carnot engine running between the same baths. The Carnot efficiency becomes a universal limit value for the efficiency $\eta$ of any engine, and hence we write

$$\eta = \frac{W}{Q_H} = \frac{Q_H - Q_C}{Q_H} \leq 1 - \frac{T_C}{T_H}. \tag{1.2}$$

Here, $Q_H$ and $Q_C$ are the heat received and rejected in the engine operation and $W$ is the net work produced in the engine cycle, as illustrated in Figure 1.

We remark that Carnot bound, like all the thermodynamical laws, strictly belongs to the classical realm of macroscopic objects. In this short review, we will describe the recent research to examine if Carnot wrote his law on a stone, or if the promised quantum technologies can operate with efficiencies exceeding Carnot's limit. Remarkably, whenever we can associate a modern device, such as masers, with a heat engine, it seems we cannot escape the Carnot bound. On the other hand, this seems to happen when the device is designed to harvest classical resources. Hence the total system (the device and the resources) is not fully quantum but a semi-classical model. Naturally, researchers ask if an entirely quantum scenario, where even the resource (or fuel) is a quantum object, should obey the Carnot bound or not?

This short review focuses on the description of efforts to reveal limits on quantum thermal machines designed to harness quantum information, more specifically quantum coherence. We shall present the achievements on the definition of heat and work applicable to such devices, including heat and work equivalent of quantum coherence, and the research summary of generalization of thermodynamical laws to the quantum realm. In particular, we follow the footsteps of Sadi Carnot and explore the bounds on the motive power of quantum information. It is the flow of heat from a hot reservoir to the cold one that produces coherent (useful) motion or motive power. Hence, it is natural to consider information flows to examine their motive power. In other words, the quantum information content of a given state should change, ideally wholly erased, to assess its useful work or heat value. For that aim, information engines that operate between non-thermal, information-containing, quantum reservoirs are suitable settings. Such a strategy allows us to establish the efficiency of such information fueled engines compared to the Carnot bound. The basic idea is to consider the source and the sink in Figure 1 as information reservoirs and to use a quantum system as the working substance of the engine. The work output of the cycle becomes then analog of a calculation or a computational process. Accordingly, whenever the information reservoirs can have heat and work equivalent, then the engine can have a classical analog cycle for which Carnot bound would be unavoidable. On the other hand modifications of the Carnot limit, proper association of "temperature" for a non-thermal, information reservoir, re-definitions of work and heat as quantum objects will be necessary, as pointed out by pioneering works in the field of quantum thermodynamics [6–9]. The term quantum thermodynamics may not sound very meaningful at first sight, as it suggests merging two theories designed for widely different scales of microscopical and macroscopical physical realms. On the other hand, Planck's law of blackbody radiation [10] and Einstein's concept of photon [11] are the historical examples demonstrating the critical role played by the thermodynamical laws in the foundations of quantum mechanics [12]. Attempts to derive thermodynamical laws from quantum mechanical principles are then naturally followed [6–9, 13, 14]. We will briefly introduce how the concepts of temperature, heat, and work enter the quantum description of modern devices, thanks to the systematic studies of the rapidly emerging field of quantum thermodynamics. We refer to more comprehensive reviews and books for a much more expanded and broader view on the subject of quantum thermodynamics [6–9, 14–17] and limit ourselves here to the perspective of quantum machines harvesting non-thermal baths, in particular quantum information resources.

Despite its abundance and cost efficiency, heat is not the main driving power of our civilization, relying on electric power. The reason is that long-time preservation of electrical resources is much easier due to the existence of an excellent range of electrical resistivity of materials compared to their thermal resistivity differences. Using quantum information as a resource for machines, either (quantum) computers, communication devices, or mechanical engines faces the same challenge. Can we keep quantum information reservoirs long enough, and can we transport quantum information to sufficiently long distances for practical purposes? We will present some potentially promising schemes from the literature to this undoubtedly a key question for the fate of quantum technologies.

A historic and paradigmatic example that a modern machine can have heat engine equivalent is the maser. In 1955, Charles Townes explained that his scientific breakthrough, the invention of maser [3, 4]. It harnesses electromagnetic energy, and it is a non-thermal device, a forerunner of the laser, which is the critical device in communication and information age. A couple of years later, Scovil and DuBois showed that this exciting, non-thermal machine is not that different than a steam engine, and hence it cannot escape from the macroscopic thermodynamical efficiency bound [2, 18]. Investigation of modern quantum information machines [9] can also follow what Scovil and DuBois did for the maser. New maser-like quantum heat engines (QHEs) can be proposed [19]. When the working substance of a machine is a quantum physical system subject to quantum mechanical transformations, or so-called quantum thermodynamic processes [20, 21], then such a device is called as a QHE [22–24]. A fundamental question in quantum thermodynamics is if and how quantumness of these heat engines can outperform their classical analogs [25]. For this short review, we distinguish those QHEs that can process classical thermal resources as semi-classical QHEs from those that operate with quantum information reservoirs, which we call Quantum Information Heat Engines (QIHEs). This term allows us to distinguish such machines form those that directly harvest quantum information irrespective if it is heat equivalent or not; such devices are called as genuine quantum information engines (QIEs) [26]. Examination of QHEs with quantum working substance harnessing classical thermal resources can be considered as a "semi-classical QHEs". Our classification is not unique, and there can be other categorizations of quantum machines [27]. Besides the type of resources, how they access the resource is another broad classification of heat engines. Depending on their continuous or periodic coupling to the baths the engines can be named as continuous or discrete (reciprocating or piston) QHEs. Continuous engines are of practical appeal due to their autonomous (self-contained) operation [23]. A remarkable result in quantum thermodynamics is that all engine types in the quantum regime of small action with respect to Planck's constant are thermodynamically equivalent [28].

Semiclassical QHEs with single atom [29], single electron [30] and single spin [31], as well as nanomechanical machines that can harvest non-classical (such as squeezed thermal bath) [32] resources have been experimentally demonstrated. Role of quantum coherence in the working system to transport energy more efficiently has been experimentally explored in a superconducting qubit simulator of biological light harvesting complexes in photosynthesis [33]. It is experimentally demonstrated that quantum coherence between internal levels in nitrogen-vacancy center semiclassical QHE can boost the power beyond classically attainable values, being first observation of a quantum advantage signature in a heat machine [34]. In addition to the working system and the baths, a battery or work reservoir (load) can be integrated to the engine system to make it more practical, as demonstrated for a single qubit working system with a load of a vibrational mode [35].

## 2. Zeroth Law of Quantum Thermodynamics

The first step in the formulation of the laws of classical thermodynamics is the introduction of the temperature concept. While the concepts of temperature and thermal equilibrium, as well as their relations, were known before, the term "zeroth law of thermodynamics" for the existence of temperature as the central postulate was first proposed by Fowler and Guggenheim in 1939 [36]. Specifically, they wrote "If two assemblies are each in thermal equilibrium with a third assembly, they are in thermal equilibrium with each other". Thermal equilibrium among several assemblies is conditioned by the equality of a specific single-valued function T, called "temperature". The temperature function depends on the thermodynamic state of the assemblies. Anyone of the assemblies can be used as a "thermometer", to determine the temperature on a suitable scale. In addition to absolute temperature scale, an "empirical temperature", can be used by choosing arbitrary scales. An example is to measure the volume of a system with a fixed quantity at constant pressure and use the measured volume as an empirical temperature.

In quantum mechanics, the temperature is not associated with a quantum operator. It is not observable, though effective or virtual temperatures are found to be convenient to introduce [37]. Indirect, operational definitions, such as spin temperature, can be defined [38]. Even in the presence of quantum coherence, useful definitions of temperature, such as apparent temperature, for interpretation of energy flows can be proposed [39]. Temperature definition for a single quantum object, i.e., without using an ensemble approach, is far from trivial [40–44]. Accordingly, a direct generalization of the classical zeroth law to the quantum realm is not available. Even the simplest quantum systems, such as a two-level atom, are not thermodynamically trivial. A single thermodynamical variable dual to energy cannot represent them unless they carry no quantum coherence in their energy basis. One can use the ratio of the excited and ground-level populations to assign a temperature to a two-level atom for example. However, if there is a quantum coherence between the two levels, it can also be used as an energy resource. Accordingly, the atom requires at least two variables dual to energy to describe its thermodynamic value. Despite the challenges of temperature definition in quantum systems, there is much demand from quantum technologies to develop temperature measurements for small quantum machines, and quantum thermometry is an essential part of quantum metrology [45].

In contrast to many works on the first and second laws of quantum thermodynamics, attempts to formulate the zeroth law remain relatively rare [6, 46–48]. Resource theory [49, 50] is used to generalizing the transient nature of equilibrium and equivalence relation form of the zeroth law to the quantum regime. In terms of an equivalence relation, the zeroth law states that if $A \sim B$ and $B \sim C$, then $A \sim C$. Here, $A, B$, and $C$ refer to quantum states, and the temperature parameter emerges as a label to distinguish different equivalence classes. In quantum thermodynamics single (temperature) parameter is not sufficient for tagging equivalence classes, and classical form of the zeroth law can be violated in the presence of quantum coherence and correlations [51]. Zeroth law, as an equivalence relation, is used to build the first and the second laws defining the operations on classes and monotonicity of generalized free energies. Equilibrium states, as canonical thermal Gibbs states, are considered as allowed free states in resource theory. Accordingly, zeroth law defines the set of available free states for further quantum thermodynamic operations. Temperature can be introduced as follows. Two thermal states with respect to their corresponding Hamiltonians are equivalent resources provided that no work can be extracted from the joint state of their arbitrary number of copies of them [52].

Quantum thermodynamics allows for both thermal and non-thermal baths [53] of finite, small, sizes. In the case of a quantum system exchanging energy and entropy with another quantum system, concepts of equilibrium and temperature should be revised. Zeroth law can also be proposed as a necessary and sufficient equilibrium condition $F(\rho_A) = 0$ [51]. Here, $F$ is the free energy and $A$ represents a set of non-interacting quantum systems $A_i$ with $i = 1..n$ with Hamiltonians $H_i$. Free energy is defined as the difference between the mean internal energy $E(\rho) = \text{Tr}(H\rho)$ and the bound energy $B(\rho)$. Bound energy is the amount of energy that cannot be extracted further by entropy preserving operations, while entropy preserving operations can access the internal energy. In this picture, zeroth law can be seen as a consequence of information conservation, while the first and second laws bring additional energy conservation constraints to the thermodynamic processes. Due to the ensemble averaging and expectation values used in the definitions, this form of zeroth law does not apply to single-shot or single-copy settings where there are no copies of the system or the environment. In equilibrium, the total system and the local systems are in their respective completely passive states [54, 55] that can be put in Gibbsian forms $\rho_A = \exp\left(-\beta \sum_i H_i\right)/\text{Tr}(\exp\left(-\beta \sum_i H_i\right))$ and $\rho_i = \exp(-\beta H_i)/\text{Tr}(\exp(-\beta H_i))$, respectively. Here the common $\beta = 1/k_B T$ is the inverse temperature and $k_B$ is the Boltzman constant; $T$ is dubbed as "intrinsic temperature". Intrinsic temperature generalizes the temperature concept to non-thermal states that cannot be written in Gibbsian forms and hence useful for finite size non-thermal reservoirs. Application of intrinsic temperature concept to the case of a system of two bosonic modes where one mode acts as a finite size bath to the other has been discussed quite recently [56].

From the dynamical description of thermalization of an open quantum system, subject to a heat bath at temperature $T$, zeroth law can be stated as the existence of the stationary state with a unique canonical thermal (Gibbs) expression at $T$ [57]. The existence of such a stationary Gibbs state is constrained, however, as it depends on the structure of the system-bath interaction. Furthermore, it depends on the open system dynamical equation, required to be in the Lindblad-Gorini-Kossakowski-Sudarshan (LGKS) [58–60]. Moreover, it requires the lack of any integrals of motion for the system [57, 61, 62]. LGKS form emerges under the Born and Markov approximations, which ensure that the reservoir state can be taken as frozen in time, and the bath has no memory so that the evolution is local in time, respectively [63, 64]. The definition of temperature for an open two-level system has been discussed very recently [65]. Non-Markovian effects on the definition of temperature in open quantum systems have also been considered [66]. Thermalization and equilibration have also been discussed in closed quantum systems by introducing a generalized notion of Gibbs (maximum entropy) ensemble, quantum typicality [67], non-integrability [68], and so-called eigenvalue thermalization hypothesis [69].

## 3. First Law of Quantum Thermodynamics

First law of classical thermodynamics is expressed as

$$dU = dW + dQ. \tag{3.1}$$

Here đ is the inexact differential representing process-dependent infinitesimal changes. It clearly distinguishes two ways of changing internal energy ($U$) of a system in terms of doing work ($W$) on it or by injecting the heat into the system ($Q$). The contributions of coherent (work) and incoherent (heat) energy transfers are additive, and the first law can be taken as the definition of heat as well. On the other hand, such a clear identification of work and heat definitions for a quantum system is unknown [70]. Earliest proposal uses a mathematical distribution of the exact differential in an infinitesimal change of the internal energy of a quantum system described by Hamiltonian $H$ and in the state (density matrix) $\rho$ [71]

$$dU = d\text{Tr}(\rho H) = \text{Tr}(H d\rho) + \text{Tr}(\rho dH), \tag{3.2}$$

and takes the first and second terms as the definitions of heat and work, respectively, such that

$$dQ := \text{Tr}(H d\rho), \tag{3.3}$$

$$dW := \text{Tr}(\rho dH). \tag{3.4}$$

Ad hoc introduction of inexact differential definitions of work and heat is an ambiguous classification of energy processes for quantum systems. One could heat (or cool) a quantum system without using thermal reservoirs, for example, by applying coherent pulses and interactions sequentially to atoms. The physical and microscopic nature of heat in an already microscopic quantum system is not explicit after this definition. Energetic resource value of quantum coherence and correlations, or quantum information in general, may challenge this classification as being incomplete, too. In particular when the system attached to the heat and work reseroirs can have quantum coherence and correlations among its constituents, the question of heat and work equivalent of these coherence and correlations both in the bath [72] and in the system should be carefully addressed before using these defitions.

In terms of the energy eigenvalues $E_i$ of $H$, and populations of the corresponding eigenstates $p_i$, the work and heat definitions become [73]

$$dQ := \sum_i E_i dp_i, \tag{3.5}$$

$$dW := \sum_i p_i dE_i. \tag{3.6}$$

These definitions are intuitively appealing as they state that processes associated with population changes of energy levels transfer heat into the system; while those processes that vary energy levels without changing their populations move energy into the system in the form of work. They form the basis of direct generalization of classical thermodynamical processes, as well as quantum heat engine cycles to their quantum counterparts [20, 21]. We remark that one can envision quantum heat machines without classical counterparts, such as algorithmic quantum engine cycles, that allows for approaching energy extraction scheme as a quantum computational method [74].

Dynamical expressions for the heat and work are proposed by examination of the master equation of an open quantum system coupled to reservoirs which gives [71]

$$Q(t) \quad := \quad \int_0^t \mathrm{d}\tau \, \mathrm{Tr} \left\{ \frac{\mathrm{d}\rho(\tau)}{\mathrm{d}\tau} [H_0 + V(\tau)] \right\}, \tag{3.7}$$

$$W(t) \quad := \quad \int_0^t \mathrm{d}\tau \, \mathrm{Tr} \left\{ \rho(\tau) \frac{\mathrm{d}V(\tau)}{\mathrm{d}\tau} \right\}, \tag{3.8}$$

where $V(\tau)$ represents the potential describing the influence of external agents on the quantum system, such as compressing the volume of confining box or electromagnetic coherent drives. The bare, unperturbed Hamiltonian of the free quantum system is denoted by $H_0$. The time derivative of $\rho(t)$ can be replaced by the dissipator superoperators of the corresponding reservoirs in the master equation [71]. The dynamical first law is expressed by $U(t) = Q(t) + W(t)$, where

$$U(t) := \mathrm{Tr} \left\{ \rho(t) [H_0 + V(t)] \right\}. \tag{3.9}$$

A related concept is the "ergotropy" which quantifies the maximum amount of work that can be harvested from non-passive states using optimized unitaries $V(t)$ [75]. Coupling of the quantum system to the external field results in work which can be probed by optical or vibrational spectroscopic methods [76]. It is straightforward to express the dynamical first law in terms of time derivatives of $E_i$ and $p_i$ [77].

There is a perspective of quantum thermodynamics that can be considered as generalization of stochastic thermodynamics so that quantum extensions of Jarzynski equality or Tasaki-Crooks relation can be established [15, 78–80], work and entropy changes can be treated as random variables so that these definitions only determine the first moment or mean of general work and heat distributions. A quantum work definition based upon Bohmian framework, consistent with positive work distribution in terms of trajectories in phase space, has been recently developed [81]. Due to sensitivity of work output to fluctuations in initial conditions and interactions, such work distributions are critical to characterize fluctuations in the performance of QHEs.

Resource theoretic approaches to the first law have been proposed as well [82]. It is shown that transformation of a system from a state $\rho$ to $\sigma$ has to satisfy the condition [82]

$$\alpha \Delta W_1 + \beta \Delta W_2 = E_B(\rho) - E_B(\sigma). \tag{3.10}$$

Here, a global system is considered that consists of a primary system, a so-called bank system (for example, a thermal reservoir at a temperature T), and two so-called batteries that store the energy (or work). An agent can make any state transformation by utilizing the bank and the batteries. The bank exchanges different amount of resources $\Delta W_1$ and $\Delta W_2$ (defined for each battery, as the difference in the monotones for the initial and final battery states) at "exchange rates" ($\alpha$ and $\beta$), to the changes in a specific monotone, quantifying the corresponding resource, $E_B$ between $\rho$ and $\sigma$. More explicitly, $E_B$ stands for the relative entropy distance of the relevant state to the bank states. A monotone can be defined in terms of the relative entropy distance $D(\rho||\sigma) = \text{Tr}(\rho \log \rho - \rho \log \sigma)$ such that

$$E_F(\rho) = \inf_{\sigma \in F} D(\rho||\sigma). \tag{3.11}$$

Thermodynamic analog of the monotone would be the free energy, while $\Delta W_1 = -\Delta U$ and $\Delta W_2 = \Delta S$ would be the internal energy and the entropy. Together with $\alpha = 1/T$ and $\beta = 1$, one recovers the usual first law of thermodynamics. The first law of thermodynamics can be established as a time-translation symmetry principle and mutual effects of quantum coherence and thermal operations can be discussed from such a perspective [83].

A complete implementation of quantum machines contains measurement stages, which contribute to the first law as energetic costs. Accordingly, quantum measurement generalizations to the first law are significant for quantum technologies, and it has been discussed in literature [84]. As a related perspective, proposals of operational definitions of work quantifiers can be noted [85, 86]. A combination of operational approach and collision model allows for the introduction of work and heat concepts at a single event level [87, 88]. Optimization of the number and kind of thermal operations subject to a set of second laws of quantum thermodynamics for work harvesting is of fundamental and practical interest [89].

## 4. Second law of quantum thermodynamics

The second law of quantum thermodynamics, similar to its classical counterpart, imposes constraints on the quantum thermodynamical processes. Laws of classical thermodynamics are of phenomenological character, and especially for the second law of thermodynamics, mathematically rigorous and conceptually transparent derivation efforts, starting from microscopical, quantum statistical, principles are rare [90]. Remarkably, different forms of second law can be challenged and extended within classical thermodynamics entirely, without any quantum generalization [91]. One can also ask if classical thermodynamics put any restrictions on quantum theory [92]. The classical second law can be expressed in terms of free energy inequality. A generalization of this approach by introducing a family of free energies and constraints on quantum state transformations by quantum thermodynamical operations has been proposed [52]. Its limiting case coincides with the classical second law, showing satisfactorily that classical thermodynamics can be envisioned as the limiting form of a more general quantum thermodynamical theory. Family of second laws is expressed in terms of inequalities of generalized free energies

$$F_\alpha(\rho, \rho_T) := kT D_\alpha(\rho||\rho_T) - kT \log Z, \tag{4.1}$$

that depends on Rényi divergences $D_\alpha(\rho||\rho_T)$, which can be related to so-called Rényi entropies for a universal approach to quantum termodynamical formulation [93]. Such an approach can be significant in the broad perspective of quantum cybernetics, whose objective is to develop a regulator to perform the desired state transformation, maintenance, and manipulation [94]. Here, the regulator can be a thermal or non-thermal environment, and the quantum thermodynamic second laws can set the bounds on the successful state transformations.

Resource theory has been applied to nonequilibrium thermodynamics and information theory connection regularly [95]. Resource theory formulation of second law has been generalized to quantum regime by faithfully taking into account the resource value of quantum correlations, in particular entanglement and coherence [96]. Quantum coherence and correlations require a cost of production as they are resources [97, 98], which should be adequately translated to the first and second laws of quantum thermodynamics in the case of quantum coherent or correlated systems [99]. Even catalytic use of coherence has been argued to be limited fundamentally [100, 101]. By identifying quantum thermal operations and their resource theoretic descriptions, fundamental bounds on quantum information-thermal engines have been provided [102]. Geometrical approaches allow to development of theoretical limits on equilibrium entropies as well as work [103].

Stochastic thermodynamics and fluctuation relations can be generalized to the quantum regime. This yields extension of the Jarzynski fluctuation theorem to the quantum case [104]. Treating work and heat as random variables, second-law type conditions on higher moments, beyond usual mean value laws, can be found and called as additional energy-information relations (AEIRs) [105]. The second law, both in classical and quantum versions, impose irreversibility conditions on state transformations [106]. One can approach to this problem by looking for generalizations of nanoscale fluctuation theorems to the quantum realm [107–110]. It fundamentally establishes practical consequences on the efficiency of quantum thermal or information machines, and significant for the realization of quantum technologies.

More specific information-theoretic expressions for quantum thermodynamical second laws can be given as a result of data processing inequality, which becomes equivalent to the statement of positivity of entropy production $\sigma = \Delta S - Q/T$ [111–113], with $S = -\text{Tr}(\rho \log \rho)$ being the von Neumann entropy. Relation of information-theoretic second-laws to the mutual information is significant for quantum technology applications such as quantum error correction [114] and minimal energetic costs of information processing [115].

Correlations and coherence in small quantum systems, or general quantum information, play a significant role in proper formulation of the second law of quantum thermodynamics [116, 117]. Fundamental bounds on work extraction from quantum states, particularly quantum coherence, can have practical applications in quantum information engines [118, 119]. Transformation of systems from an initial to a target non-thermal (non-equilibrium) state under regulating feedback and measurements are constrained by generalized second laws involving quantum information contributions [120–124]. It is noted that problem of quantum measurement and quantum operational definition of work are closely related [125]. Quantum information contribution to the classical second law can be given in terms of quantum-classical mutual information $I$ term in the following inequality [126, 127]

$$W_{\text{ext}} \leq -F + k_B T I,  \tag{4.2}$$

Quantum information plays a central role in quantum thermodynamics [7, 15, 112]. Generalized second laws, in the form of generalized engine efficiency bounds in the presence of quantum information or memory effects, have been proposed [128]. Such bounds can surpass the classical Carnot limit as can be seen from the expression [128]

$$\eta := \frac{W_{\text{ext}}}{Q_H} = 1 - \frac{T_L}{T_H} + \frac{k_B T_L (\Delta S - \Delta I)}{Q_H}.  \tag{4.3}$$

This particular example requires that imperfectly closed cycle, in which the memory is not reset to its initial state. Associated expression of second law in terms of free energy is found to be [128]

$$W_{\text{ext}} \leq -\Delta F_S + k_B T C,  \tag{4.4}$$

where $C$ depends on initial quantum discord. Information contribution to the second law allows for bypassing the requirement of having two thermal bath for work extraction so that one can envision harvesting useful work out of single heat reservoir [129] or single quantum system [130], and furthermore it is possible to surpass the classical Carnot bound [131].

A dynamical approach to the second law of quantum thermodynamics requires a proper derivation of quantum open system dynamics [132–135]. Time-dependent formulations of the second law of quantum thermodynamics involve the definition of heat flows [136] that can be determined with thermodynamic consistency under Davies construction [58] out of the Gorini-Kossakowski-Lindblad-Sudarshan (GKLS) quantum master equation [59, 60]. Defining the heat current

$$J(t) := \mathrm{Tr}\left(H(t)\frac{d\rho(t)}{dt}\right) = \mathrm{Tr}(H(t)\mathcal{L}(t)\rho(t)), \tag{4.5}$$

where $H(t)$ is the Hamiltonian of the system and $\mathcal{L}(t)\rho(t)$ is the superoperator describing the dynamical influence of the bath coupled to the system in state $\rho(t)$. Time-dependence of the $H(t)$ is slow enough to satisfy the quantum adiabatic theorem. This assumption can be relaxed and more general master equation can be derived as shown recently [137]. Consistent with the zeroth law, we have $\mathcal{L}(t)\rho_G(t) = 0$ for instantaneous Gibbs states, $\rho_G(t) = \exp\left(-H(t)/kT\right)/Z(t)$ with $Z(t)$ being the instantaneous partition function. The first law, for the case of multiple baths with corresponding dissipation generators $\mathcal{L}_i(t)\rho(t)$ then reads as

$$\frac{dU(t)}{dt} = \sum_i J_i + \mathrm{Tr}\left(\rho(t)\frac{dH(t)}{dt}\right). \tag{4.6}$$

The second term represents the power absorbed by the system, while the first term is the net current injected into the system from the respective baths. The first law can be generalized to the case of particle reservoirs [138], as well as to the case of information or quantum coherence reservoirs [139]. The second law is expressed as the positivity of the global system's entropy production rate, consisting of the main system and the baths, [6, 13, 71]

$$\frac{dS(t)}{dt} - \sum_i \frac{J_i}{k_B T_i} \geq 0. \tag{4.7}$$

This Clausius form of the second law of quantum thermodynamics is consistent with the H theorem and rigorously derived after the Spohn inequality [13] applied to the baths under detailed balance (Kubo-Martin-Schwinger) condition [140, 141]. Here, $S(t)$ is the von Neumann entropy of the system.

Reciprocity relations, known as Onsager theorem [142], among the energy and particle flows, can be stated as the fourth law of thermodynamics. Their generalization to the quantum regime in the presence of quantum coherent reservoirs has been recently introduced [139]. Investigating the role of quantum coherence in the system or the reservoirs to improve the performance of a heat engine and set new efficiency limits as generalized Carnot's version of the second law, play a central role, both fundamentally and practically, in quantum thermodynamics (for a recent review, see Ref. [143]).

## 5. Third law of quantum thermodynamics

According to the third law of thermodynamics, a system cannot be cooled down to absolute zero by any finite-time (or finite number of steps) process. It is established by the Nernst theorem that sets vanishing entropy change in the limit of zero temperature [144, 145]. Several works are focusing on placing the third law to quantum thermodynamics perspective [146–151]. Currently, the number of of studies of the quantum version of the third law is relatively low. Remarkably, fast refrigeration using reservoirs, which are compact, small, readily integrable to quantum devices, can have significant impact on quantum technologies, such as quantum computers in practice [152]. Early focus on the first and second laws of quantum thermodynamics could shift towards the third law in near future. An intriguing relevant direction of research here is to use control theory to accelerate the thermalization times [137], inspired by the success of shortcuts designed to shorten the time of quantum adiabatic transformations [153–159], which has been experimentally demonstrated [160–162].

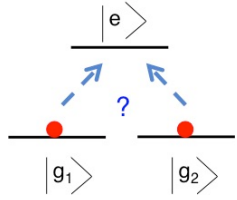

Figure 2: A beam of three-level atoms used to drive a micromaser; each atom in the pump beam is assumed in a coherent superposition state of lower two levels $|g_1\rangle$ and $|g_2\rangle$.

The other promising direction is to consider non-thermal, information, reservoirs for fast and compact cooling possibilities. In the rest of this short review, we will focus on quantum coherent reservoirs, their scaling and quantum advantages, and their general classification for heat and work processes.

## 6. Quantum coherence as a Maxwell demon

In 2002, a pioneering study showed that classical Otto bound can be improved by a quantum Otto engine [163]. It is found that one can obtain laser action in the hot exhaust gases of a heat engine, which can be envisioned as a "quantum afterburner". Even though the machine is still limited by the Carnot bound, it provided significant impetus to look for quantum advantages to improve classical devices' efficiencies using their quantum analogs. It is, however, not expected that one can enhance the effectiveness of a Carnot engine. Carnot bound is a form of the second law of thermodynamics, and even though it is stated for classical macroscopic systems, there is a central belief among scientists that it is scaling invariant.

However, a dramatic proposal suggested that while the second law can still be valid, a new limit that can be higher than the classical Carnot bound can be imposed on the efficiency of a QHE [164]. Specifically, the authors consider a quantum photonic Carnot engine in a micromaser setting. The enhancement is coming from a profound quantum effect known as quantum coherence. The pump atoms, driving the micromaser cavity, are ejected out of a heated oven and injected a small bit of quantum coherence. Each pump atom is taken as a three-level system, as shown in Figure 2, where the lower pair of levels are in a superposition (coherent) state $|\Psi\rangle$. Quantum coherence subtly plays the role of a Maxwell demon that allows for extracting work from a single thermal reservoir.

The unnormalized state $|\Psi\rangle$ can be written as

$$|\Psi\rangle \sim |g_1\rangle + e^{i\theta}|g_2\rangle \qquad (6.1)$$

$\theta$ is the phase associated with the atomic coherence. As a new control parameter, the phase can be varied to increase the radiation field's temperature and to extract work from a single heat bath. Considering that some light is sending on to this atom, the atom absorbs the light and gets excited. After absorbing the light, the state $|\Psi\rangle$ becomes $|\Psi\rangle \rightarrow (1+e^{i\theta})|e\rangle$ (cf. Figure 2). Excitation can be canceled by taking $\theta = \pi$, while de-excitation is allowed; which is a situation similar to Young's double slit interferometer. Such an atom can efficiently transfer energy to a resonator. When the coherent atoms form a beam, we can envision two reservoirs interacting with a micromaser cavity. One is of the atoms at their lower levels so that it can be assigned zero temperature; another one is those atoms at their excited levels that can be associated with negative zero temperature [165]. If the atoms carry quantum coherence, their interaction with the cavity can be turned off and turned on depending on which state they are injected into the cavity. Detailed balance and the thermal equilibrium will still be established, but the cavity temperature now can be controlled, and enhanced, by the quantum coherence. At the specific quantum coherence with $\theta = \pi$, atoms injected to the micromaser cavity at lower levels will not absorb energy from the cavity electromagnetic field. Those atoms injected at their excited states, on the other hand, will release energy into the field. Using quantum coherence, we have an energy sorting machine that is what the Maxwell demon would do. The second law is protected as such a demon is internal to the system, not an external agent tracking every atom. In the historical example, the demon would be the gatekeeper separating hot and cold particles; while here every atom has an internal demon that tells them effectively either to go through the cavity or not.

In Ref. [164], a three-level single atom absorbs some energy from the heat bath with temperature $T_H$ then goes through the cavity where they release some of their energy to the cavity field. There is also a cold bath (entropy sink) with temperature $T_C$. If the atoms have no quantum coherence, the engine's efficiency will be the same as the Carnot efficiency $\eta = 1 - T_C/T_H$. With the injected coherence, the efficiency $\eta_\theta$ depend on the initial superposition angle $\theta$ as

$$\eta_\theta = \eta - \pi \cos\theta, \tag{6.2}$$

which overcomes the Classical Carnot bound at $\theta = \pi$, setting a larger limit for the performance of the QHE. For equal temperatures of cold and heat baths, $T_C = T_H$, classical efficiency will be zero while the QHE can still work with the efficiency $\eta_\theta = -\pi\cos\theta$ and produce work from a single heat bath via trading quantum coherence.

## 6.1. Heat and Temperature value of quantum coherence

In the case of equal temperatures for heat and cold bath that we mentioned above, the heat engine can be illustrated in Figure 3a with injected some coherence to the initial state. The device would act like a perpetual motion machine of the second kind, if we ignore the role of quantum coherence injection for its operation. In such a perpetual motion machine, one can harvest energy with perfect efficiency; no energy ever leaks away as exhaust heat, and the device can operate forever on fixed energy supply, violating Kelvin's statement of the second law of thermodynamics that states, no process is possible whose sole result is the absorption of heat from a reservoir and complete conversion of this heat into work. On the other hand, injection of quantum coherence introduces a second bath with elevated effective temperature, higher than $T$, so that the machine in fact operates between two reservoirs without any violation of the second law, as illustrated in Figure 3b.

Though the machine obeys the second law, its performance can still be beyond the limits of classical Carnot bound. In Figure 3b, an engine system is in a single environment, but the heat intake is modified by quantum coherence. Coherence effectively elevates the environment temperature during the heat injection stage of the engine. After injecting coherence to a bath at temperature $T$, the bath essentially becomes a non-thermal environment. If one can design a thermodynamic engine cycle in which the working system could perceive the

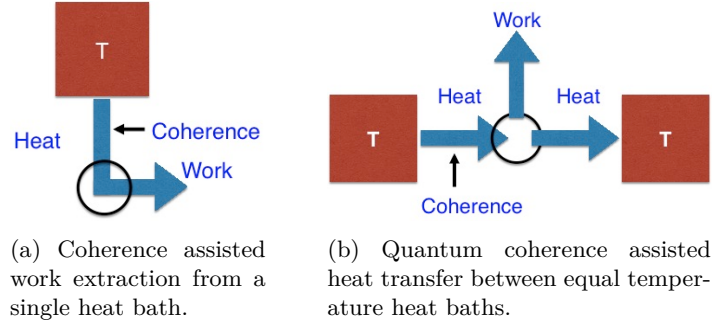

(a) Coherence assisted work extraction from a single heat bath.

(b) Quantum coherence assisted heat transfer between equal temperature heat baths.

Figure 3: Quantum coherence can allow for thermodynamical processes forbidden by classical thermodynamics. Careful examination of quantum coherence assisted processes reveal that such cases are not in conflict with the second law of thermodynamics, though generalization or enhancement of thermodynamical efficiency limits can be achieved.

non-thermal environment as an effective heat bath at an effective elevated temperature, such a temperature would be a genuine temperature as far as the working system is considered [27, 72]. Such an interpretation is consistent with an operational definition of the temperature of a single quantum object, here the working system. One can assign the coherence dependent temperature $T(\theta) > T$ to the non-thermal bath and treat it as an effective hot bath. Accordingly, such a device can have formally the same Carnot bound expression, though it is higher than the classical one

$$\eta_\theta = \eta - \frac{T}{T(\theta)} \ . \tag{6.3}$$

and the QHE can perform a task not available to any classical engine that it can produce work using a single heat bath.

## 6.2. Quantum coherence as a Maxwell demon in collision model thermalization

Collision model is a natural route to thermalization (see Ref. [139] and references therein) and can be used in more general framework of open quantum systems and nonequilibrium dynamics [87, 88, 166]. According to the Rayleigh model of thermalization, a massive object (thermometer, for example) collides with tiny particles (projectiles or fuel atoms) in the same thermal state randomly and exchange energy, an equilibrium will eventually be established. The massive object will reach a thermal state at the same temperature as the projectiles. But what if these projectiles would not be in thermal states but some special non-thermal, quantum coherence, state [72, 139]. Moreover, could we take a tiny, quantum object, such as an atom or a cavity field, to replace the massive object, as our quantum thermometer. Could the thermometer object still be found in a thermal state? The answer depends on the type of quantum coherence in the non-thermal state. For certain quantum coherences, the thermalization of the massive object is still possible. Such coherences basically change the collision rate of the projectiles with the massive object and acts as another temperature knob to thermalize the thermometer to different temperatures.

An anology can be drawn to the idea of Maxwell to introduce a demon who can sort the hot and cold molecules in an ensemble to alter the temperature as illustrated in Figure 4. In this picture, the right (R) box denote the thermometer system, the circles with different filling colors indicate fuel atoms in lower and excited states. The quantum coherence carried by the fuel atoms alter the energy emission and absorption rates with the thermometer (denoted by $p_e$ and $p_g$), that can be envisoned as coherence controlled collision rates. Figure 5a and 5b shows an example where the thermometer is taken as an optical cavity and a two-level atom, respectively. If the thermometer is found to be in a thermal state (determined from the population distribution of the energy levels as shown in Figure 5c) at a temperature $T$ after random interactions with the non-thermal atoms, we can assign $T$ as an effective temperature of the non-thermal bath. More explicitly we can use

$$\frac{p_e}{p_g} = e^{-\hbar\omega/k_B T}.$$ (6.4)

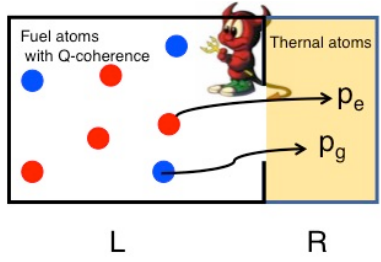

Figure 4: Selection probabilities (or collision rates indicated by $p_e$ and $p_g$) of the fuel atoms in the left (L) part of the box, colliding with the thermometer system, represented by the right (R) portion of the box, are modified by quantum coherence (Maxwell demon sorting). The box is assumed to be always in a thermal state, and any atom passes into the right box from the left will be in a thermal state, without any coherence.

These populations, $p_e, p_g$ can be envisioned as collision rates of hot (excited) and cold (grounded) atoms with the thermometer system. To see this explicitly, we assume one interaction of the fuel and thermometer atom at a time, and each collision happens at a random time and takes a short time relative to natural time scales of thermometer and fuel atoms. Under these assumptions, we can derive a master equation for the system from which we get

$$\dot{\rho}_{ee} = -p_g \rho_{ee} + p_e \rho_{gg},$$ (6.5)

$$\dot{\rho}_{eg} = (i\omega - (p_g - p_e)/2)\rho_{eg}.$$ (6.6)

Here, $\rho_{ee}$ and $\rho_{ee}$ are the excited and ground-state populations in terms of density matrix diagonal elements in the thermometer atom energy basis. $\rho_{eg}$ is the off-diagonal element (coherence) of the density matrix of the thermometer atom. An ensemble of atoms all in an excited state can be taken as a bath at temperature $-0^o$C while an ensemble of atoms in the ground state would correspond to a bath at $0^o$C, as depicted in Figure 6. If we solve the master equation for the steady-state, we get the canonical Gibb's state,

$$\rho_{ee}^{ss} = \frac{e^{-\hbar\omega/k_B T}}{1 + e^{-\hbar\omega/k_B T}},$$ (6.7)

$$\rho_{eg}^{ss} = 0;$$ (6.8)

defining a temperature for the thermometer atom same as in equation (6.4). The thermometer temperature can be controlled by the populations of the excited and ground states; while the populations (or the collision rates) could be controlled by the quantum coherence in the fuel atoms. For that aim, we consider multilevel and multiatom quantum coherent systems.

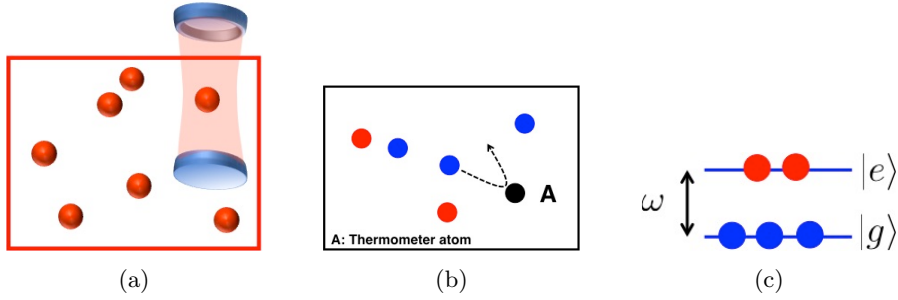

(a)             (b)             (c)

Figure 5: (a) A quantum thermometer (cavity) is placed in a non-thermal bath of atoms with quantum coherence. (b) A thermometer atom collides with an atom of a non-thermal bath randomly at a time. Collision rates of A with red (hot) and blue (cold) atom are $p_e$ and $p_g$ respectively. (c) A thermal atom has more population in the ground state and the less population in the upper state. Ground and excited states are represented by $|g\rangle$ and $|e\rangle$, respectively, and $\omega$ is the frequency difference between them.

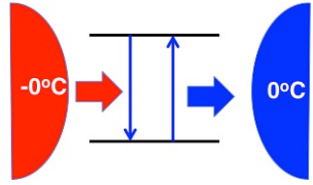

Figure 6: Excited state is on the left and ground state is on the right. Their temperatures are at $-0^oC$ and $0^oC$, respectively.

### 6.2.1. Operational work and heat definitions for non-thermal states

We consider an information-carrying object $R$ is a fuel system that can be harvested as heat by a system (thermometer) $S$ with Hamiltonian $H_S$. Let a global unitary $U$ act on the composite system $\rho_{RS} \equiv \rho_R \otimes \rho_S$. We can introduce the energy difference through the changes of the state of $S$ by a map

$$\rho_S \rightarrow \rho_S' = \text{Tr}_R(U(t,0)\rho_R \otimes \rho_S U(t,0)^\dagger). \tag{6.9}$$

where the reduced states of $R$ and $S$ are given by $\rho_R = \text{Tr}_S(\rho_{RS})$ and $\rho_S = \text{Tr}_R(\rho_{RS})$. We define an "operational heat" based on the properties of $\rho_S'$. We remark that while one could write $\rho_S' =: \Lambda(t,0)\rho_S$, to define a dynamical map $\Lambda(t,0)$, such a map is not necessarily a quantum dynamical semi-group in general. We do not restrict the general definition of operational heat and work to the Markovian dynamics. A natural condition to define heat is that $\rho_S'$ to be formally written in a canonical Gibbs state, which is a classical-like state such that it is diagonal in the energy basis with eigenvalues decreasing with energy. If the thermometer state is initially in a Gibbs form and remains Gibbsian, then the map is called a Gibbs preserving map [167–169], in which case we assume the $R$ is a heat-like resource. An effective temperature can be assigned to $R$. The energy transfer $\delta U$ from $R$ to $S$ can be identified as operational heat $\delta Q$ [170], which reads after a single collision as

$$\delta U = \delta Q = \text{Tr}(H_S \rho_S') - \text{Tr}(H_S \rho_S). \tag{6.10}$$

On the other hand, $R$ is considered a work-like resource if $S$ is injected coherence after the collision. To classify quantum states as heat-like or work-like, it is necessary for us to find out which quantum states can lead to Gibbs preserving changes in the thermometer state.

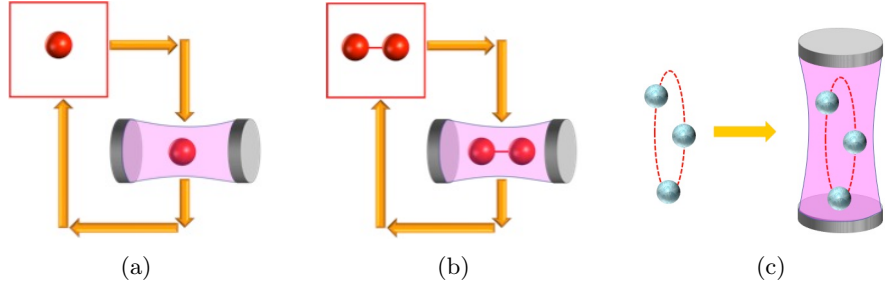

Figure 7: (a) Two-level atom and cavity system. After the interaction, certain coherence will be injected to the cavity state and it will be non-thermal. (b) Two-atoms in a certain coherent state that go through the cavity can preserve the Gibbs state of the cavity and change its temperature. (c) 3-atom cluster in certain entangled states can preserve the cavity field in thermal state and change the cavity temperature more efficiently than the two-atom fuel.

In what follows, we will only focus on $N$-qubit cluster quantum fuels. Though, similar properties are expected for a so-called $N$-level phaseonium (generalized two-level atom with $N$-fold degenerate lower levels), we will not consider it here. A detailed comparison and advantages of cluster fuel relative to phaseonium fuel, together with exploration of quantum entanglement as a refined heat exchange coherent fuel, can be found in Ref. [171].

It is recognized that if $R$ is a two-level atom, then quantum coherence in it destroys the thermalization [72]. Accordingly, coherent two-level atom state is a work-like coherence. As an example, a two-level atom ($R$) fuel and cavity thermometer system $S$ is illustrated in Figure 7a. In contrast, one can find a two-atom fuel with a specific coherence in its energy basis that preserves the Gibbs state after a collision, which is illustrated in Figure 7b. Different temperatures can be induced on the cavity field depending on the coherence between $|eg\rangle$ and $|ge\rangle$ states [171]. For example,

$$|\psi^+\rangle = \frac{1}{\sqrt{2}}(|01\rangle + |10\rangle), \tag{6.11}$$

$$|\psi^-\rangle = \frac{1}{\sqrt{2}}(|01\rangle - |10\rangle). \tag{6.12}$$

these Bell states induce a thermal state in the quantum thermometer at different temperatures $T_\pm$. Compared with a two-atom fuel in the mixed state, the hierarchy of the temperatures is found to be

$$T_+ > T_{\text{mix}} > T_- \equiv T_{\text{env}}. \tag{6.13}$$

The symmetric Bell state induces higher temperatures on the thermometer relative to the mixture and anti-symmetric Bell state. These results can be generalized to three-qubit fuel states (see Figure 7c) and a general classification of quantum coherences for their heat and work equivalence can be obtained [72]. Note that these works consider many collisions instead of a single collision. The dynamical map is purely Markovian without any memory effect at all times and the collision model has no explicit time dependence. Accordingly, a micromaser type coarse-grained master equation of the Lindblad form can be developed. Its steady state is used for classification of quantum coherences as heat exchange, displacement and squeezing coherences [72]. Ref. [170] presents a "microscopical surgery" of this approach and generalizes the definitions to the single collision level.

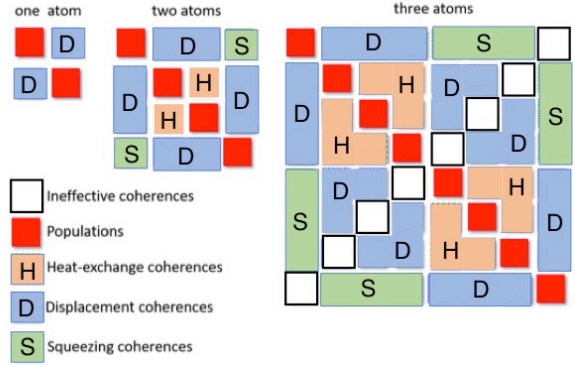

Figure 8: (Reproduced from Ref. [72]) Density matrix in energy basis for different number of atoms. Regions of coherences with different thermodynamic functionality are denoted by $D$, $H$, and $S$, which refers to displacement, heat-exchange, and squeezing coherences. The white squares are ineffective coherences.

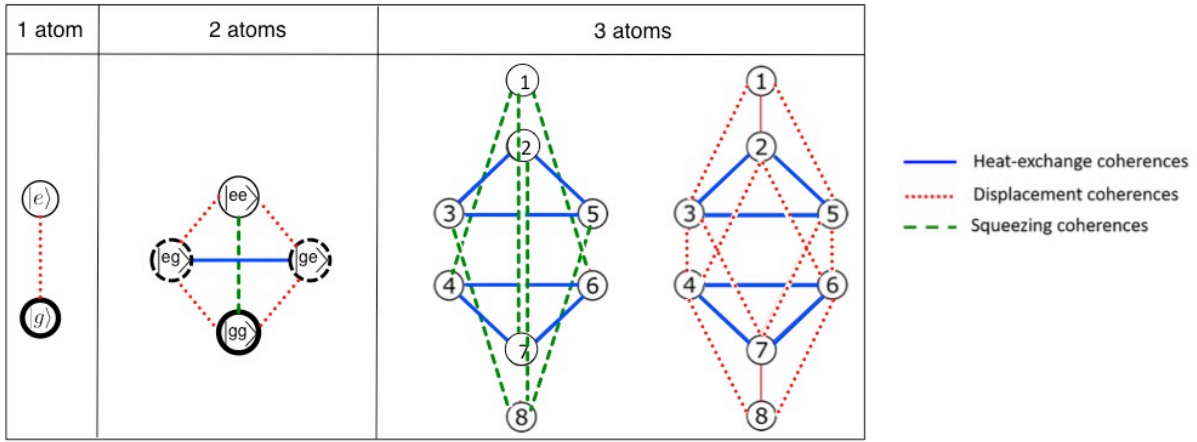

Figure 9: Graphical (tree-like) representation of quantum coherences in the energy basis of one-, two- and three-atom clusters are given from left to right. The circles denote the energy basis states, where we distinguished the different number of excited atoms by thin-solid and thick-solid and dashed lines. The numbers in circles mean the computational basis for the three-atom cluster. Blue-solid links connecting 2, 3, and 5 (and 4, 6, and 7) indicate heat-exchange coherences. Displacement coherences (red dotted lines) are between states differing by one excitation, and squeezing coherences (green dashed lines) are between states differing by two excitations (illustrated separately for the 3-atom cluster).

Including only one coherence in two-atom fuel can be associated with a regular Maxwell demon with two arms making energy sorting or changing collision rates with cold and hot atoms. Such a sorting process could be more efficient if the demon had more arms. This could be accomplished by considering larger atomic clusters [72]. It is found that atomic fuel clusters in Dicke-like or $W$-type states would yield Gibbs preserving maps and allow for efficient temperature control of the thermometer system. Density matrix structure for a multi-atom cluster is given in Figure 8, where $D$, $H$, $S$ denotes displacement, heat-exchange, and squeezing coherences, which are diagramatically defined in Figure 9. The main diagonal denotes the populations. The heat-exchange coherences (HEC's) are those that are populating the main diagonal blocks. They yield Gibbsian thermal states in the thermometer. Quantum machines harvesting HECs then becomes subject to Carnot's law with an Carnot limit enhanced by HECs the baths. There are also ineffective coherence terms that do not contribute to the field evolution by short-time collisions with a thermometer atom [172]. Work-like coherences, displacement coherences are those that yield non-thermal coherent atomic states in the thermometer atom. Squeezing coherences lead to squeezed state if the thermometer system is cavity field. They could contribute to the populations and boost the temperature if the thermometer is a single two-level atom. However, squeezing coherences cannot contribute unless displacement coherences present, and hence they effectively lead to non-Gibbsian states, too. Accordingly they are not classified as HECs. HECs can be converted to heat, and heat [173–175] can produce them.

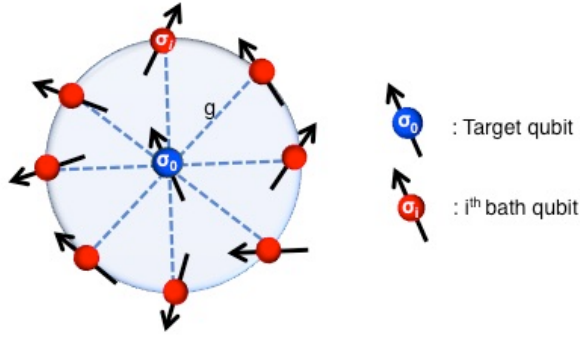

Figure 10: Interaction of a single (target) qubit with a $N$-qubit atomic cluster (quantum fuel). The coupling $g$ of the target qubit to each identical bath qubit (spin-1/2 particle), labeled with $i = 1, 2, ..., N$, is supposed to be same. After each interaction, the state of the quantum fuel is reset.

The problem of thermalization of a composite system consisting of a target qubit (the system) that randomly and repeatedly interacts with a cluster of N identical spin-1/2 (qubit) particles is examined in Ref. [172, 176]. The system is illustrated in Figure 10. The evolution of the target qubit corresponds to a coherently driven two-level atom in an effective squeezed thermal bath. The participation of squeezing, coherence, and thermal components to the overall system are disjoint from each other, and they can be controlled, turned on or off, by choosing corresponding quantum coherences in the cluster state.

There is significant scaling advantage of quantum fuels based upon HECs as the number of coherences grow up rapidly, with the number of atoms $N$ [172],

$$N_{\text{HEC}} = \binom{2N}{N} - 2^N. \tag{6.14}$$

HECs contribute to the temperature through their additive contribution to the heat transfer rates $p_e$ and $p_g$. In terms of collective spin of the atomic fuel cluster we can write $p_e = \langle J_+ J_- \rangle$ and $p_g = \langle J_- J_+ \rangle$. These expressions reveal that $p_e$ is the sum of HECs and populations lying above the anti-diagonal of the density matrix o the fuel cluster in energy basis. Similarly $p_g$ is determined by the sum of the populations and HECs below the antidiagonal. While the number of coherences grow faster than exponential (after $N = 4$), translation of scaling advantage to temperature is far from trivial due to weak coherences and possible scaling up of quantum dephasing and decoherence effects due to exposure of the large clusters to larger environments. When HECs are used as quantum fuels, one can use larger coherences as HECs yield genuine Gibbs states in thermometer system. A critical size of the fuel cluster can be determined for given scaling of the environmental decoherence. Optimization of the bath cluster size, together with the scaling advantages in the working system [177], and finally establishing a network of size optimized engine-bath systems can be envisioned for high performance, high power quantum energy machine networks or quantum machine factories.

Classification of quantum coherences in an $N$-qubit cluster [72, 172] or $N+1$-level (Two-level atom with $N$-fold degenerate ground state) atom [178] depends on the structure of the interaction between the target system and the fuel cluster. The classification is given for the target system either a two-level atom or a single mode cavity; and the interaction is limited to single excitation exchange (Tavis-Cummings model or dipolar interaction). In such a case, HECs are those between the degenerate energy levels while the displacement and squeezing coherences are between the levels separated by single or double excitation frequencies. In literature, HECs are also identified as "horizontal" coherences, while displacement or squeezed coherences are called as "vertical" coherences [143, 179]. Role of HECs in thermodynamics processes involving heat flows have been discussed, in particular in controlling direction of heat flow beyond classical constraints [39, 139]. In other studies on heat flow management by quantum coherence, HECs are called as "internal coherences" or "non-energetic coherences" [180]. Fast equilibration times, enhancement of work output and efficiency can be scaled with superlinear laws using heat exchange coherences [172], which sometimes can also be identified as "bath-induced coherences" [143]. They are proposed to be advantageous for quantum thermometry, too [181].

Further applications and advantages of HECs for QHEs will be presented in the following sections. If the thermometer or working system is a composite object with several subsystems, then quantum coherence among them can play a role on heat flows and on engine performance, too. Furthermore, in a genuine QHE with quantum resources and work systems, quantum coherence can exist between the bath and the working system, as well. Such cases are beyond the scope of present review (see for example Ref. [143] and references therein).

The different coherences and their relation to the number of excitations are illustrated in Figure 9. The vertical coherences (between the states in different energy levels) are work-like, and horizontal coherences (between the states in the same energy level) are heat-like. The atoms carrying heat-exchange or horizontal coherences can be considered as quantum fuel molecules that can power QHEs. The vertical or displacement/squeezing coherences can power so-called quantum thermo-mechanical machines [27].

In such a composite system, asymptotic thermalization or its absence is strongly dependent on the initial state of the $N$-qubit cluster [172]. The target qubit evolves into a thermal state if the cluster carries only heat exchange coherences. Quantum thermalization of the target qubit with such a cluster exhibits substantial quantum advantages in thermalization time and accessible temperature scaled by $N^2$ [172].

## 7. Applications and Advantages of Quantum Fuels

### 7.1. Collective scaling advantage in quantum fuels for work enhancement

The typical application of quantum thermodynamics is QHEs. However, in contrast to their classical analogs, such tiny engines produce practically negligible work and power; accordingly, a natural question is how their power can be enhanced. For that aim, without increasing the size and complexity of the quantum working system, one can take advantage of scaling quantum fuels. As an early example, it is proposed that a photonic QHE work output can be improved by using superradiance when pumped by multi-qubit coherent clusters [182]. It is showed that the work output depends on the number of the atoms quadratically due to the superradiance [183], where a cluster can radiate quadratically faster than a single atom. The model is illustrated in Figure 11. Superradiance (SR) is given as a cooperative emission of light from an ensemble of excited two-level atoms in a small volume relative to emission wavelength [183]. The atoms radiate in a synchronized (coherent) manner, at a quadratic rate with the number of atoms. Also experimental verification of SR has been achieved in wide variety of systems, including atomic gases, quantum dots, and atomic Bose-Einstein condensates [184–191].

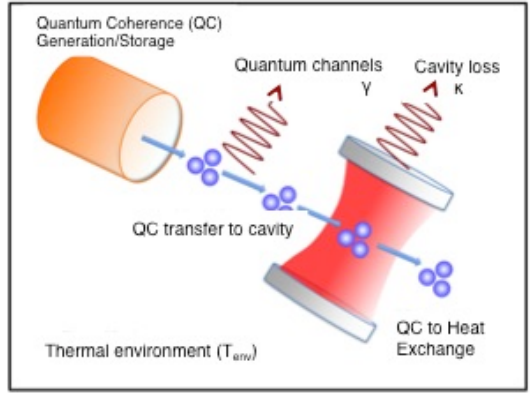

Figure 11: A cavity field pumped by a beam of $N$-atom clusters in atomic coherent or Dicke states reaches a steady-state, a coherent thermal state. The corresponding work output of the QHE can be determined from the steady-state photon number, which is enhanced by the superradiance, scaled with $N^2$.

## 7.2. Quantum advantages and quantum law of diminishing returns

It is challenging to isolate quantum fuels of large atomic clusters from the decoherence effects of the environment. In fact, even for a single three-level atom with quantum coherent lower levels (phaseonium quantum fuel) subject to dephasing makes the realization of photonic QHEs that can harvest quantum coherence too challenging [192]. The $N^2$ advantage in the work output enhancement can be severely degraded or even prohibited due to a similarly scaled increase in the environmental decoherence. In Ref. [178], a photonic Carnot engine harvesting a quantum fuel of a beam of multi-level coherent atoms is considered under constant, linear, and quadratic scaling models of decoherence. It is found that there is a critical size of the cluster for which quadratic enhancement in the work output can still be obtained even in the presence of a quadratic increase in the decoherence factor. We can represent coherences between $N$ atomic energy levels as a complete graph as in Figure 12, which can be envisioned as a quantum information-fuel molecule. According to the result of Ref. [178], there is a critical size of such a quantum fuel molecule, after which further increase of the size would give negative yields, rapidly decreasing the efficiency and work output.

Work output ($W$) and thermodynamic efficiency ($\eta$) are calculated for different decoherence factors, $\zeta = e^{-xt}$, $\zeta = e^{-Nxt}$, and $\zeta = e^{-N^2xt}$ where $x$ is the atomic decay rate, $t$ is exposure time to the environment, and $N$ denotes the number of coherent quantum levels. Dependence of $W$ and $\eta$ has a general form,

$$\eta, \ W \sim N^2\zeta(N), \tag{7.1}$$

which can be compared to the microeconomical law of diminishing returns [193], plotted in Figure 13.

### 7.2.1. Thermal charging and discharging of quantum coherent fuel clusters

The finite size systems with a varying number of particles are further examined in [173] from the perspective of their generation using thermal and non-thermal means. The system subject to thermal charging of coherences is illustrated in Figure 14a. A pair of two-level atoms is initially prepared in a state with an amount of heat exchange coherence $C_L$. Using a thermal bath collectively coupled to the particles, the initial state can be transformed into another state with higher coherence $C_H > C_L$. The role of dipolar interactions between the particles is shown to be not critical in quantum induction of coherence. The discharge of stored coherences as heat is shown in Figure 14b, where a single two-level atom collides sequentially with a pair of atoms with coherence $C_H$. Some amount of the energy cost to generate coherences in the initial state can be harvested back as heat. Eventually, the single-atom reaches a steady-state that is a thermal state with an effective temperature of $T_{\text{eff}}$. This temperature depends on the coherence $C_H$ of the sub-environment atomic pairs as illustrated in Figure 14c.

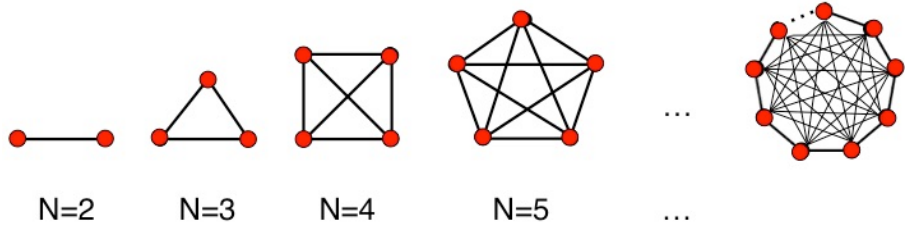

Figure 12: Representation of coherences between $N$ atomic energy levels in a multilevel atom or an atomic cluster as a complete graph. The graph can be envisioned as a quantum fuel molecule of size $N$ such that larger $N$ causes more exposure to the environment, causing more rapid decoherence characterized by a decoherence factor $\zeta(N)$. Different models of $\zeta(N)$ have been explored, and it is found that there is a critical size of quantum fuel for which quantum advantages in the enhancement of work and efficiency can still be found in the presence of decoherence [178].

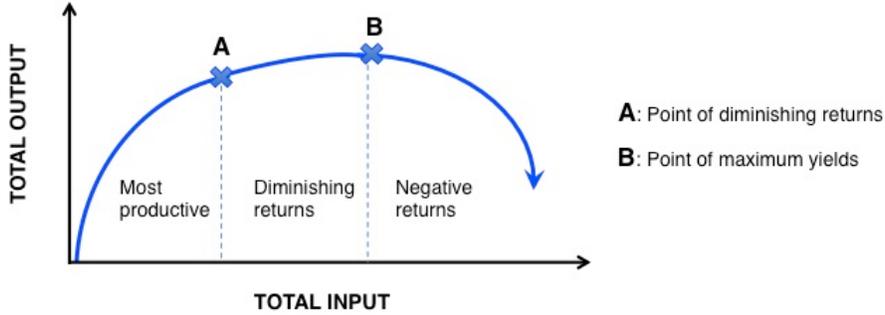

Figure 13: Microeconomical law of diminishing returns. Advantages of using more input resources, such as workers, is most effective to some critical point A. After that, the returns diminish till to point B, where the output is maximum. After B, output decreases with further increase of input, and the returns become negative.

### 7.3. Remote superradiant heating with a spin-star quantum fuel

An $N$-element spin-star network, illustrated in Figure 15, has been proposed as a quantum fuel to power up a photonic Carnot engine [176]. Previously, it has been shown that using individual qubit in a pair yields different work output than using the pair as whole as quantum fuel [194]. When the network is in thermal equilibrium with a heat bath, the outer spins are in a non-equilibrium state, and the central spin is in an effective thermal state. The effective temperature is higher than the bath, and it exhibits super linear scaling with $N$. Also, the nonlinearity can be adjusted to $N^2$, $N^3$, or $N^4$ with the coupling's anisotropy parameter. The central-spin beam effectively acts as a hot reservoir to the cavity field and brings it to a thermal steady state. The efficiency is found to be several orders of magnitude higher than typical PCEs. It is also discussed in [176], a similar scheme where a single macroscopic spin replaces the outer spins. It has been proposed that such single-particle quantum thermalizing devices could be realized using superconducing circuit QED schemes [195].

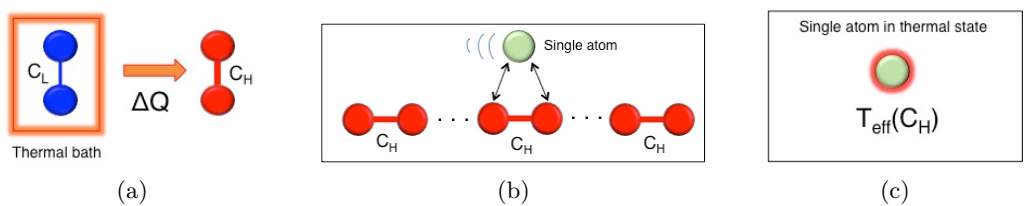

(a)　　　　　　　　　　(b)　　　　　　　　　　(c)

Figure 14: (Reproduced from Ref. [173])A pair of two-level atoms initially in a state with (lower) coherence $C_L$ can be transformed into another state with higher coherence $C_H > C_L$ by collective interaction with a thermal bath. Heat is used to produce ("charge") coherences. (b) A single two-level atom is coupled sequentially with a pair of atoms with coherence $C_H$. Some amount of dissipated energy in order to generate coherence in initial state can be harvested back by a repeated interaction method. The coupling happens at random times. Coherences are harvested back ("discharged") as heat. (c) The single atom reaches a steady state eventually that can be described by a thermal state with an effective temperature $T_{eff}$ that can be controlled by the $C_H$ of the sub environment atomic pairs.

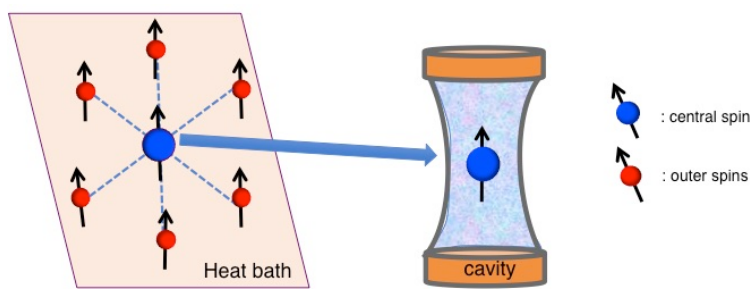

Figure 15: A spin star system that consists of a central spin coupled to $N$ surrounding spins. The spin star system is in thermal equilibrium with a heat bath at temperature $T$. An optical cavity is held at the same temperature in the same environment. Central spin has a local temperature higher than the heat bath. If a beam of such central spins are injected into a cavity then the cavity can be equilibrate at a higher temperature than the environment. The enhancement in the cavity temperature is proportional to $N^2$ or higher powers. If the setup is operated in a quantum heat engine, such scaling advantage can be translated into the work output and efficiency of the machine. The effect is dubbed as "remote superradiance", since the collective ensemble does not immersed into the cavity as a whole. The effect is shown to be related to quantum correlations in the spin-star quantum fuel of the engine.

## 8. Conclusion

This brief review first presented a summary of efforts on extending the four laws of classical thermodynamics to the quantum domain. Second, we focused on the description of studies of non-thermal baths considered to power up quantum heat engines. Specifically, a broad classification of quantum coherences for their work and heat value, together with their applications for quantum thermodynamical tasks, are reviewed. Rapidly evolving literature on the enhanced Carnot limit in the presence of such non-thermal quantum coherent baths is introduced. Studies on the identification of heat-equivalent quantum coherences as quantum fuels, which can be produced by heat and converted into heat, are shortly described. As a result of these efforts, depending on the isolation and storage of quantum coherences, it can be envisioned that quantum machines can be empowered with compact, safe, rapidly and naturally chargeable, clean quantum coherent information fuels and they can exhibit superior efficiency than their classical counterparts.

Further significance of quantum coherent information molecules can be expected in quantum sensing and metrology [196–200]. Using QHEs for thermometry [201] and magnetometry [202] has been proposed. Other potential impact of quantum fuels can be envisioned in the fields of quantum batteries [203, 204], and thermal quantum annealing [204–206] or error correction [114]. In addition, it can be a promising direction to explore global coherences as resources for topological quantum heat engines [207, 208]. Optimization of quantum thermodynamical processes can benefit from modern quantum machine learning methods [209]. This concise review is neither complete or exhaustive, nor intended to be a comprehensive discussion of rapidly evolving field of quantum thermodynamics. It is our hope that it would first serve as an introduction to present state of the laws of quantum thermodynamics; second, it would overview accumulated information on work and heat equivalent of quantum coherence to power up quantum heat and information engines.

## Acknowledgment

Ö. E. M. acknowledges the hospitality of TÜBİTAK Research Institute for Fundamental Sciences (TBAE) during the Workshop of Quantum Frontiers of Technology, where some parts of this review was delivered as an invited talk.